\documentclass[twocolumn,prb,aps,superscriptaddress,floatfix]{revtex4-2}

\usepackage{amsmath,amssymb}
\usepackage{graphicx}
\usepackage{bm}
\usepackage{hyperref}
\usepackage{xcolor}
\usepackage{placeins}
\usepackage{comment}
\usepackage{caption}
\makeatletter
\AtBeginDocument{%
  \long\def\@makecaption#1#2{%
    \par
    \vskip\abovecaptionskip
    \begingroup
      \small\rmfamily
      \begingroup
        \flushing
        \let\footnote\@footnotemark@gobble
        \@make@capt@title{#1}{#2}\par
      \endgroup
    \endgroup
    \vskip\belowcaptionskip
  }%
}
\usepackage{tikz}
\usepackage{pgfplots}
\pgfplotsset{compat=1.18}
\usepgfplotslibrary{groupplots}
\usetikzlibrary{arrows.meta, positioning, calc, decorations.pathmorphing,
                 backgrounds, fit, shapes.geometric, patterns, shadows.blur}

\definecolor{pumpred}{HTML}{C0392B}
\definecolor{probeblue}{HTML}{2980B9}
\definecolor{samplefill}{HTML}{D5E8D4}
\definecolor{sampleedge}{HTML}{82B366}
\definecolor{paramblue}{HTML}{DAE8FC}
\definecolor{paramedge}{HTML}{6C8EBF}
\definecolor{fwdgreen}{HTML}{D5E8D4}
\definecolor{fwdedge}{HTML}{82B366}
\definecolor{predorange}{HTML}{FFE6CC}
\definecolor{prededge}{HTML}{D6A656}
\definecolor{lossred}{HTML}{F8CECC}
\definecolor{lossedge}{HTML}{B85450}
\definecolor{adpurple}{HTML}{E1D5E7}
\definecolor{adedge}{HTML}{9673A6}
\definecolor{measbox}{HTML}{FFF2CC}
\definecolor{measedge}{HTML}{D6B656}
\definecolor{arrowcolor}{HTML}{2C3E50}
\definecolor{feedbackcolor}{HTML}{8E44AD}
\definecolor{recblue}{HTML}{2471A3}
\definecolor{truered}{HTML}{C0392B}
\definecolor{initgray}{HTML}{888888}
\definecolor{adamorange}{HTML}{E67E22}
\definecolor{forestgreen}{HTML}{27AE60}
\definecolor{activeshade}{HTML}{E8F8E8}
\definecolor{gsblue}{HTML}{BDD7EE}
\definecolor{gsedge}{HTML}{5B9BD5}
\definecolor{gsorange}{HTML}{FCE4D6}
\definecolor{gsoedge}{HTML}{ED7D31}
\definecolor{gsgreen}{HTML}{E2EFDA}
\definecolor{gsgedge}{HTML}{70AD47}
\definecolor{gspurple}{HTML}{D9D2E9}
\definecolor{gspedge}{HTML}{7E57C2}

\pgfplotsset{
    paperplot/.style={
        line width=0.6pt,
        tick style={line width=0.5pt},
        label style={font=\small},
        tick label style={font=\footnotesize},
        title style={font=\small\bfseries, at={(0.02,0.98)}, anchor=north west},
        legend style={font=\scriptsize, draw=none, fill=white, fill opacity=0.85,
                      text opacity=1, inner sep=2pt, row sep=-1pt},
        xtick pos=lower,
        ytick pos=left,
    },
}

\newcommand{\wpsq}{\omega_{\mathrm{p}}^2}
\newcommand{\wosq}{\omega_0^2}
\newcommand{\eps}{\varepsilon_0}

\begin{document}

\title{Automatic-differentiation-enabled dynamic parameter retrieval with sub-pulse-width resolution}

\author{Huaiyue Peng}
\email[]{huaiyue.peng@yale.edu}
\affiliation{Department of Physics, The Hong Kong University of Science and Technology,
Clear Water Bay, Hong Kong SAR, China}
\affiliation{Department of Applied Physics, Yale University, New Haven, Connecticut 06520, USA}

\author{Yuchen Lin}
\affiliation{Department of Physics, The Hong Kong University of Science and Technology,
Clear Water Bay, Hong Kong SAR, China}

\author{Fu Deng}
\affiliation{Department of Physics, The Hong Kong University of Science and Technology,
             Clear Water Bay, Hong Kong SAR, China}

\author{Xiaoyue Zhou}
\email[]{zhouxy@ust.hk}
\affiliation{Department of Physics, The Hong Kong University of Science and Technology,
             Clear Water Bay, Hong Kong SAR, China}

\author{Jingdi Zhang}
\email[]{jdzhang@ust.hk}
\affiliation{Department of Physics, The Hong Kong University of Science and Technology,
             Clear Water Bay, Hong Kong SAR, China}

\date{\today}

\begin{abstract}
Time-resolved terahertz time-domain spectroscopy (THz-TDS) is a phase-sensitive tool in condensed matter physics for tracking photoinduced non-equilibrium dynamics of low-energy elementary excitations. However, the measured response function, optical conductivity $\sigma(\omega,t_{pp})$, becomes unreliable in reporting the state of matter when material properties drastically change on a timescale comparable to or less than the probe pulse duration, obscuring the sub-pulse-width dynamics. To resolve this issue, we present a full-waveform inversion framework inspired by the multi-dimensional retrieval philosophy of frequency-resolved optical gating (FROG). By leveraging the automatic differentiation (AD) technique and the two-dimensional time-domain signal $E(t_{g},t_{pp})$, we show that one can uniquely solve the inverse problem, at the sub-pulse-width resolution, of retrieving physical observables that are still well-defined, i.e., time-dependent scattering rate $\gamma(t)$, plasma frequency $\omega_\mathrm{p}(t)$ and resonance frequency $\omega_0(t)$, while the response functions are not. Further optimization by gradient-based routines (Adam + L-BFGS) via JAX makes the method exceptionally robust against experimental noise and probe pulse distortions. The validity of the AD-enabled methodology is benchmarked both by a self-consistent numerical approach and by experimental data from real ultrafast THz spectroscopy measurements.
\end{abstract}

\maketitle

\section{Introduction}

Understanding elementary excitations in quantum materials stands at the forefront of contemporary condensed matter physics, which has been enriched by the advent of ultrafast optical spectroscopy from far-infrared through UV spectral range. Within this scope, terahertz (THz) radiation, which has photon energy as low as a few millielectron volts (meV), perfectly matches the energy scale of important elementary excitations in solids, e.g., free carriers and lattice modes \cite{basov2011electrodynamics, Zhang2014}. A remarkable capability of spectroscopy in the THz regime lies in its phase sensitivity, enabled by direct mapping of pulsed THz waveforms in the time domain. When the measurement is properly referenced, one can unambiguously extract both real and imaginary parts of the complex dielectric function \cite{dressel2002electrodynamics}. When paired with an excitation laser pulse, one turns the measurement into a pump-probe configuration and, therefore, gains access to dynamic information of the sample at a temporal resolution empirically limited only by the durations of the pump and probe pulses. As such, the technique, specifically the optical pump-THz probe (OPTP) spectroscopy, has proven an outstanding tool for unraveling the complex many-body interactions in quantum materials, such as high-temperature superconductors \cite{averitt2001nonequilibrium} and heavy-fermion materials \cite{zhang2018ultrafast}.

For a system in equilibrium, the optical conductivity $\sigma(\omega)$ is always well-defined and thus normally deemed an observable that can be connected with details of certain oscillator model. In a non-equilibrium but slow-varying system, the response function becomes time-dependent $\sigma(\omega,t_{pp})$ but can still be satisfactorily defined at each pump-probe delay, qualified to serve as snapshots of the dynamic system. \textcolor{black}{When fit to an oscillator model, e.g., Drude- or Lorentz-type, it renders microscopic details of the sample in terms of the scattering rate $\gamma(t_{pp})$ and the plasma frequency $\wpsq(t_{pp})$~\cite{Kindt1999,Beard2000}. However, in the presence of dynamics with a characteristic timescale faster than the finite duration of the THz probe pulse, the optical conductivity $\sigma(\omega,t_{pp})$ ceases to precisely and unambiguously track the status of the system, and it consequently fails to faithfully register the scattering rate $\gamma(t_{pp})$ and the plasma frequency $\wpsq(t_{pp})$, albeit they themselves are still definable and have physical meanings on the sub-pulse-width timescale ~\cite{averitt2000conductivity,Nienhuys2005}.} Recent work has addressed related deconvolution challenges in optical pump-probe spectroscopy~\cite{Ashoka2022}, but a general framework for extracting sub-pulse-width dynamics from OPTP data has remained unavailable.

\begin{widetext}
\begin{minipage}{\linewidth}

\includegraphics[width=\textwidth]{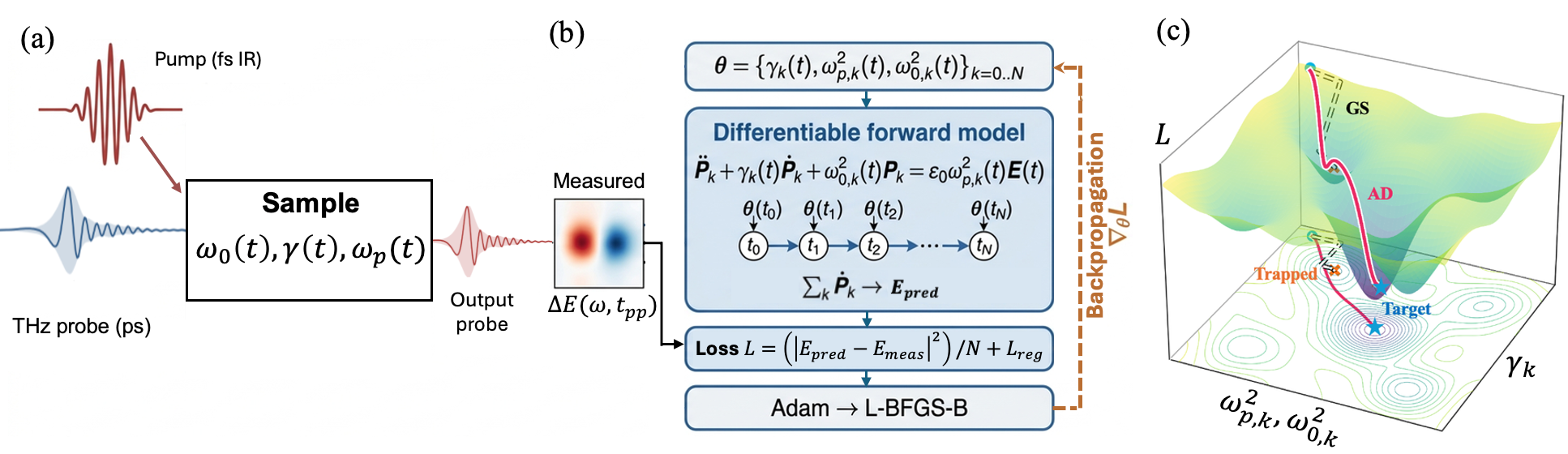}
\captionof{figure}{(a) Schematic of OPTP measurement: a femtosecond pump pulse (red) excites the sample;
a broadband THz probe with an inter-pulse delay $t_{pp}$ measures the transient state of the sample.
(b) Differentiable forward model: the equation of motion is solved step by step in \textcolor{black}{absolute time $t$},
mapping parameters $\boldsymbol{\theta}=\{\gamma_k(t),\omega_{p,k}^2(t),\omega_{0,k}^2(t)\}$
that governs the prediction on current density $J(t_g,t_{pp})$, enabling reverse-mode AD gradients.
(c) Loss landscape: AD follows exact gradients (pink) to the global minimum (blue star),
whereas gradient-free methods stagnate at local traps (orange).}
\label{fig:schematic}
\end{minipage}
\end{widetext}

To access the true $t$-dependent dynamics, one must solve a joint inverse problem \textcolor{black}{to recover the parameters definable on the sub-probe-pulse-width scale, e.g., $\gamma(t_{pp})$ and $\wpsq(t_{pp})$, which should be labeled by the absolute pump-gate delay $t=t_{pp}+t_{g}$ rather than the pump-probe delay $t_{pp}$. Note that the gate pulse refers to the optical pulse of a duration much shorter than the THz pulse, so that waveform of the latter can be obtained in an electro-optic sampling measurement by varying their relative delay ($t_g$).} 
This is mathematically analogous to the pulse-retrieval problem in frequency-resolved optical gating (FROG)~\cite{Trebino1997,Kane1999}, where one recovers the electric field of an ultrashort pulse from a two-dimensional spectrogram.
Motivated by this analogy, we first examine a Gerchberg--Saxton (GS)-like iterative projection algorithm~\cite{Gerchberg1972,Fienup1982} for retrieving $\gamma(t)$ and $\wpsq(t)$.
While this approach works for the Drude model, we find that it fundamentally fails for the Lorentz oscillator due to the increased dimensionality and the ill-conditioned nature of the projections involving the second-order equation of motion.

To overcome this limitation, we propose a different approach: we construct a fully differentiable forward model of the THz probe--sample interaction [Fig.~\ref{fig:schematic}(a)] and use automatic differentiation (AD)~\cite{Baydin2018} [Fig.~\ref{fig:schematic}(b)] to compute exact gradients of a loss function with respect to all unknown parameters.
This enables efficient gradient-based optimization using modern machine-learning infrastructure (Adam~\cite{KingmaB14} + L-BFGS~\cite{Liu1989}), without any hand-derived update rules or projection operators [Fig.~\ref{fig:schematic}(c)].
Differentiable physics simulations have recently proven powerful for inverse problems in molecular dynamics~\cite{Schoenholz2020} and fluid mechanics~\cite{Hu2020}, as well as in physics-informed neural networks~\cite{Raissi2019}; here we apply the method to ultrafast THz spectroscopy.
The AD method successfully retrieves complex, multi-parameter dynamics for both Drude and Lorentz oscillators, demonstrating its robustness against noise and broad convergence basins.
Besides numerical verification, we validate the approach with experimental OPTP data from the type-II Weyl semimetal WTe$_2$~\cite{Zhou2025}, where rich photoinduced dynamics across the $T_d$--$1T'$ structural transition provide a convincing test result.

The resulting gradient-based optimization over the full parameter trajectory, using
Adam~\cite{KingmaB14} and L-BFGS~\cite{Liu1989} requires no hand-derived update
rules. This algorithm successfully retrieves the tested trajectories for the Drude and Lorentz model, and can be extended to multi-oscillator systems~\cite{Schoenholz2020,Raissi2019}.

\section{Forward model and inverse problem}

In an OPTP experiment, two independent mechanical delay stages are utilized to vary the pump-probe delay ($t_{pp}$) and the probe-gate delay to read out the THz probe pulse ($t_g$). Under this configuration, \textcolor{black}{$t_{pp}+t_{g}$ defines the absolute time elapsed after pump excitation at the precise moment of the sampling action by the gate pulse, while $t_g$ defines the relative time between the gate and probe pulse}. \textcolor{black}{In real experiment, for each choice of pump-probe delay $t_{pp}$, one measurement on the THz probe waveform can be done by varying the relative time $t_g$ within a defined range, e.g. $t_g\in[-3, 3]$ in picoseconds}. For free carriers conforming to the Drude model, the momentum $p(t_g, t_{pp})$ obeys the equation of motion: 
\begin{equation}
    \frac{d{p(t_g, t_{pp})} }{dt_g}= -e\,E_\mathrm{probe}(t_g) - \gamma(t_{pp}+t_g)\,p(t_g, t_{pp})
    \label{eq:drude}
\end{equation}
and the associated current density is
\begin{equation}
    J(t_g,t_{pp}) = -\frac{\wpsq(t_{pp}+t_g)\,\eps}{e}\,p(t_g, t_{pp})
    \label{eq:current}
\end{equation}
where $\omega_{p}(t)$ is the time-dependent plasma frequency that reads $\omega_p^2=N(t)e^2/m^*$; $N(t)$ is \textcolor{black}{the} carrier density and $m^*$ the effective mass.
For bound charges \textcolor{black}{and} lattice modes, it comes with a finite natural frequency and follows the Lorentz oscillator model as
\begin{multline}
    \frac{d^2 r(t_g,t_{pp})}{dt_g^2}
    + \gamma(t_{pp}+t_g)\frac{dr}{dt_g}
    + \wosq(t_{pp}+t_g)r
    \\
    = -\frac{e}{m^*}E_{\mathrm{probe}}(t_g)
    \label{eq:lorentz}
\end{multline}
with $J(t_g,t_{pp}) = -\frac{\wpsq(t_{pp}+t_g)\varepsilon_0m^*}{e}\frac{dr}{dt_g}$.
In the most general cases, a multi-oscillator model can be employed to take into account the joint contribution of Drude and Lorentz oscillators, i.e., a weighted linear combination of the two.

The inverse problem may be defined as retrieving the parameter trajectory $\bm{\theta}(t)=
\{\gamma(t),\wpsq(t)\}$ (Drude) or $\{\gamma(t),\wosq(t),\wpsq(t)\}$ (Lorentz) from
the measured $\Delta E_\mathrm{meas}(t_g,t_{pp})$ at a set of $N_{t_{pp}}$ delays.
We incorporate Eqs.~\eqref{eq:drude}--\eqref{eq:lorentz} in JAX using explicit Euler
(Drude) and an implicit Verlet scheme (Lorentz) with time step $\Delta t=10$--$40\,
\mathrm{fs}$, vectorized at all delays with \texttt{jax.vmap}.
Exact gradients are computed by reverse-mode AD through the \texttt{jax.lax.scan}
time-stepping loop.
The loss function decomposes into a data-fidelity term ($\mathcal{L}_\mathrm{data}$) and a weighted sum of regularizers ($\mathcal{L}_\mathrm{reg}$), currently including temporal smoothness and positivity constraints:
\begin{align}
  \mathcal{L}(\bm{\theta})
  = \underbrace{\frac{|\textcolor{black}{E_\mathrm{rec}}-E_\mathrm{meas}|^2}
          {| E_\mathrm{meas}|^2}}_{\mathcal{L}_{\mathrm{data}}}
  + \underbrace{\sum_{k} \lambda_k\,\mathcal{L}_k(\bm{\theta})}_{\mathcal{L}_{\mathrm{reg}}}
    \label{eq:loss}
\end{align}
In regularizers $\mathcal{L}_\mathrm{reg}$, we define $\mathcal{L}_\mathrm{smooth} = \sum_\theta\langle|\nabla_t\theta|^2\rangle$ to penalize rapid temporal variations and $\mathcal{L}_\mathrm{pos} = \sum\theta\langle[\min(\theta,0)]^2\rangle$ to enforce non-negativity.
  
Optimization uses Adam ($3\,000$--$10\,000$ iterations, cosine-decay learning rate)
followed by L-BFGS refinement.
A typical run with $N_t=600$--$1800$ time points and $N_{t_{pp}}\leq 50$ delays
completes in $\sim\!1$--$5\,\mathrm{min}$ on a single CPU.

\section{Results}

\subsection{Probe-pulse-width independence}

\begin{figure}[htbp]
\includegraphics[width=\columnwidth]{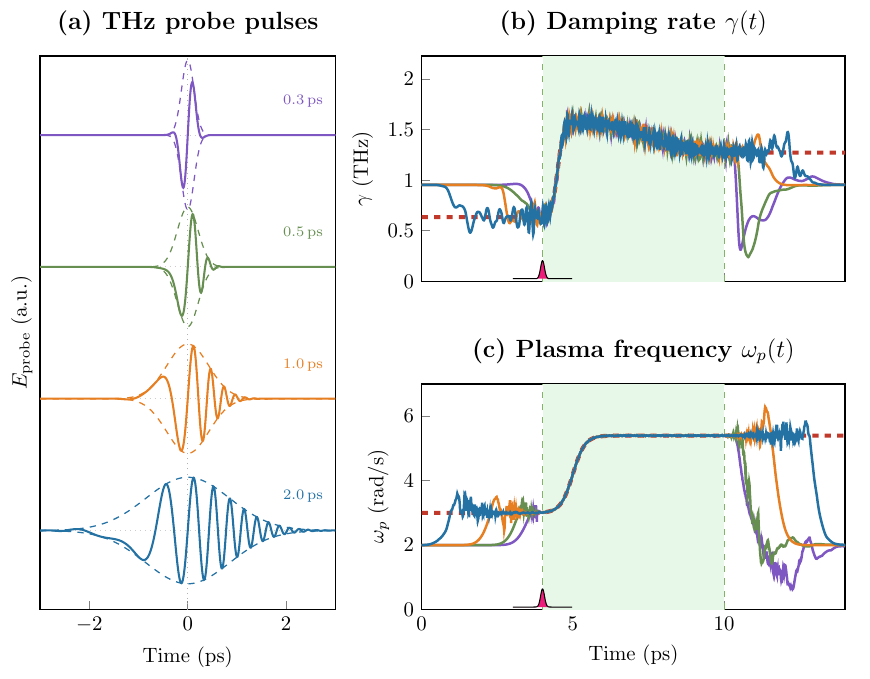}

\caption{Probe-pulse independence.
(a) THz probe pulses of various durations (carrier frequency $f_0 = 2$\, THz,
FWHM $= 0.3$, $0.5$, $1.0$ and $2.0$\,ps); dashed lines
show the Gaussian envelope and the dotted vertical line marks $t_g = 0$.
\textcolor{black}{(b-c) Time-dependent Drude parameters $\gamma(t)_{rec}$ and $\omega_p(t)_{rec}$ AD-retrieved by inverse solving with mimic data for different probe pulses (solid lines), and comparison with the ground-truth function $\gamma(t)$ and $\omega_p(t)$ (dashed lines). Line colors in panel (b) and (c) correspond to different probe durations in panel (a).} The magenta pulse schematically marks the onset of the prescribed dynamics.} 
\label{fig:probe}
\end{figure}

To validate the new methodology, we perform a consistency check with both numerically generated and experimentally measured data. With the former, we verify the independence of the reconstructed dynamics on the probe pulse duration after retrieval. We generate synthetic Drude data for a step-function $\gamma(t)$ profile
(rising from $2.0$ to $5.0\,\mathrm{ps}^{-1}$) and apply the AD method with
Gaussian probe pulses \textcolor{black}{with FWHM pulse duration 0.3, 0.5, 1.0 and 2.0 ps} [Fig.~\ref{fig:probe}]. \textcolor{black}{For the four different probe durations, one recovers the nearly identical step-wise dynamics within the time window of interest (shaded region, $4$--$10$\,ps), as defined in the standard pump-probe measurement ~\cite{Nienhuys2005}. The normalized RMS error for $\gamma(t)$ is approximately $3\%$ for all retrievals, independent of the probe pulse duration [Fig.~\ref{fig:probe}(b,c)]. The AD framework can properly retrieve, in the constrained region, the true dynamics of physical meaning ($\gamma(t)$, $\omega _p (t)$) from convoluted signals of compromised temporal resolution. In the unconstrained region, and by its name, the retrievals strongly diverge from the intended curves, but a longer probe tends to stay stable for an extended range in both time directions.} 

\subsection{Numerical verification}

\begin{figure}[htbp]
\includegraphics[width=\columnwidth]{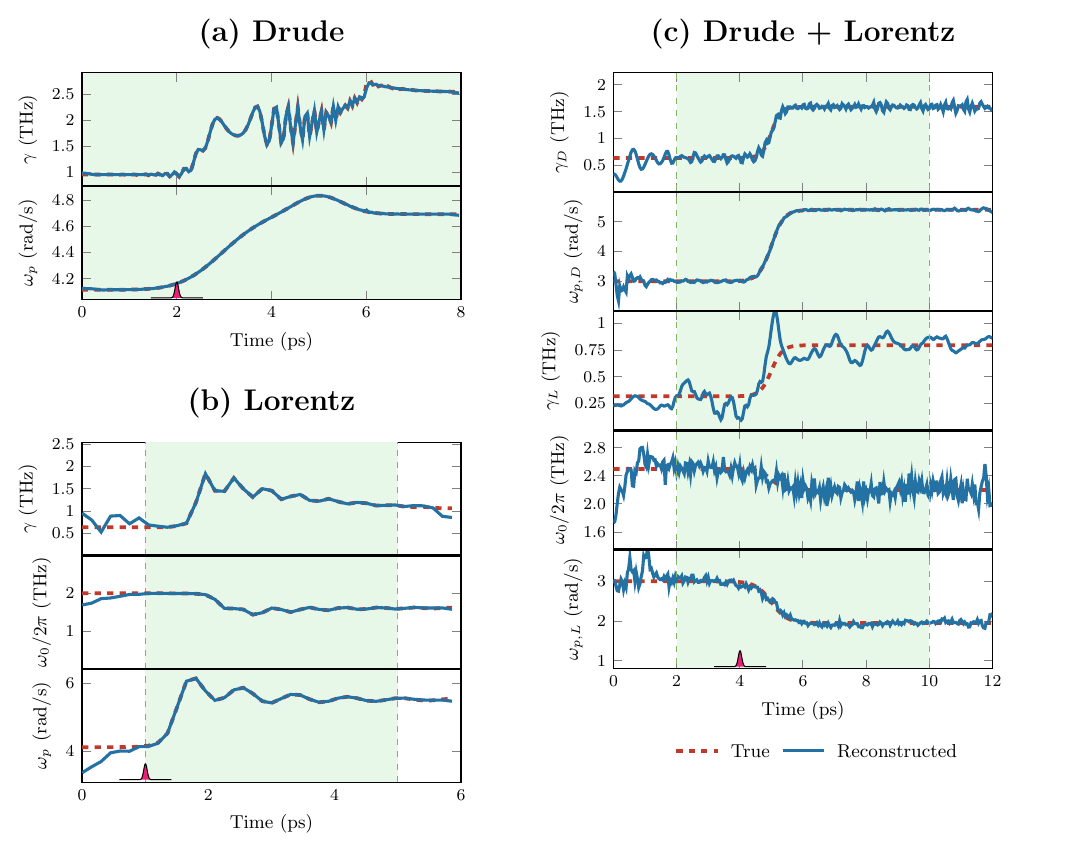}
\caption{AD parameter retrieval for three types of oscillators (red dashed line: true target dynamics; blue solid line: AD-recovered dynamics; green shade: time window of accuracy).
(a)~Drude oscillator: complex multi-scale $\gamma(t)$ and $\omega_p(t)$; probe window $t\in[0,8]$\,ps.
(b)~Lorentz oscillator: three simultaneously varying parameters $\gamma(t)$,
$\omega_0(t)/2\pi$ and $\omega_p(t)$; probe window $t\in[1,5]$\,ps.
(c)~Drude+Lorentz multi-oscillator: five parameters $\gamma_D(t)$,
$\omega_{p,D}(t)$, $\gamma_L(t)$, $\omega_0(t)/2\pi$ and $\omega_{p,L}(t)$
recovered simultaneously; probe window $t\in[2,10]$\,ps.
In all cases, the AD gradient propagates correctly through the ODE solver.}
\label{fig:models}
\end{figure}

Figure~\ref{fig:models} demonstrates the AD method in the presence of three types of oscillators as a test of its accuracy.
For the Drude model [Fig.~\ref{fig:models}(a)], we test a multi-scale $\gamma(t)$ of complex shape and a non-trivial $\wpsq(t)$; \textcolor{black}{both consist of a double-step error function and an exponentially decaying hump, but the former carries one more component –chirped oscillation in a Gaussian profile.} All features are faithfully recovered after $10\,000$ Adam + L-BFGS iterations.
For the Lorentz oscillator [Fig.~\ref{fig:models}(b)], three time-dependent parameters, $\gamma(t)$, $\wosq(t)$, and $\wpsq(t)$, are \textcolor{black}{ accurately recovered within a window of 4 ps (green shade) from a 6 ps retrieval,} demonstrating that the AD gradient propagates correctly through the implicit-Verlet
second-order ODE solver. \textcolor{black}{For the more complicated multi-oscillator system (Drude $+$ Lorentz)} [Fig.~\ref{fig:models}(c)], five parameters, $\gamma_D(t)$, $\omega_{p,D}(t)$, $\gamma_L(t)$, $\omega_{0,L}(t)$ and $\omega_{p,L}(t)$, are
independently recovered from the superimposed spectral response.
In all cases, the loss function [Eq.~\eqref{eq:loss}] and optimization framework \textcolor{black}{remain the same}, and only the forward ODE solver is modified.

\subsection{Experimental verification}
\begin{figure}[tbp]

\includegraphics[width=\columnwidth]{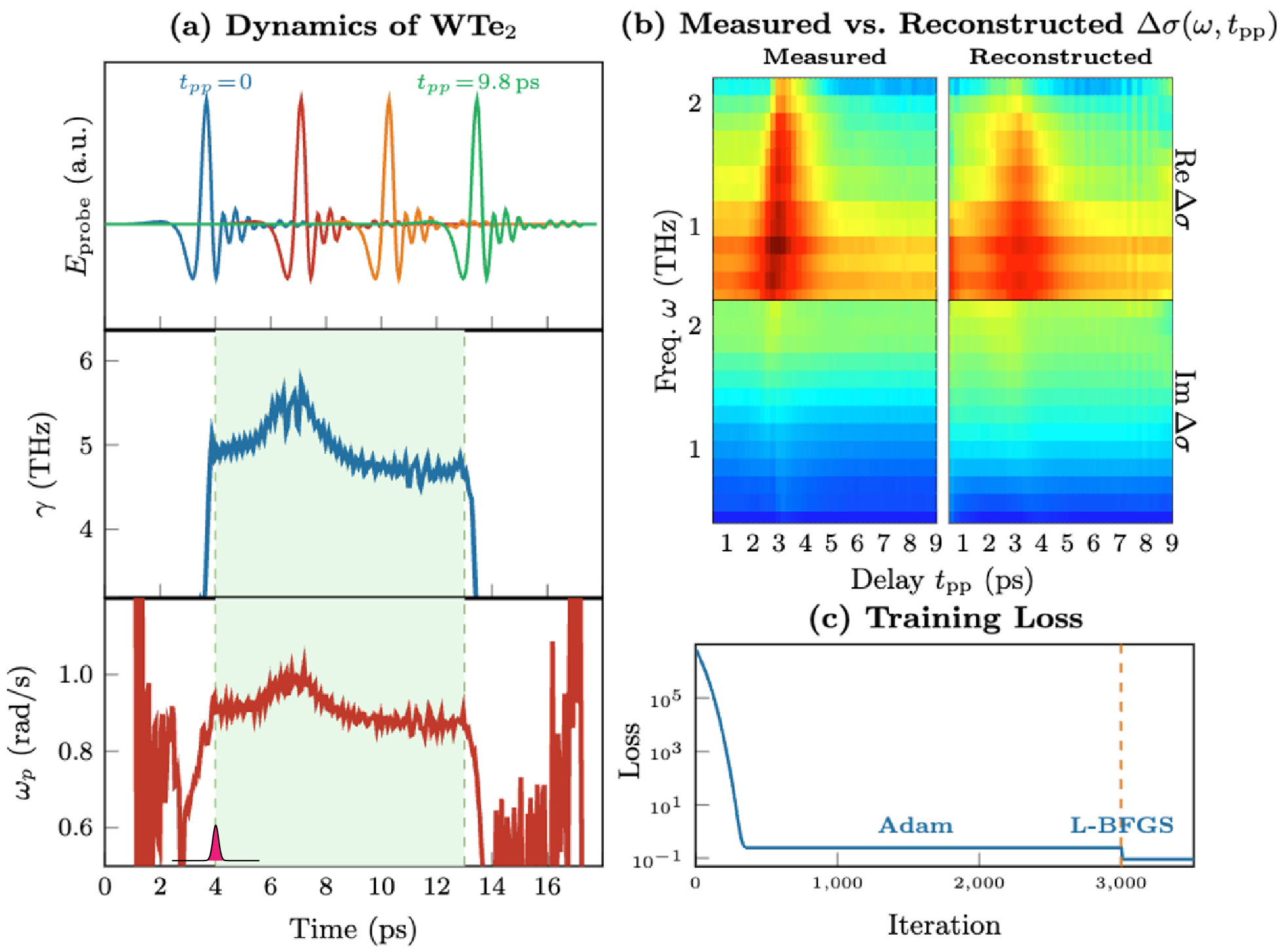}
\caption{Representative test with experimental results on WTe$_2$ thin film.
(a)~Top: terahertz probe waveforms $E_\mathrm{probe}(t_g)$ recorded at four
representative pump--probe delays (blue $\ t_{pp}\!=\!0$ to green $\ t_{pp}\!=\!9.8$\,ps).
Middle: reconstructed time-dependent scattering rate $\gamma(t)$, showing a rapid
photoinduced rise from $\!4.9\,\mathrm{THz}$ to
$\!5.4\,\mathrm{THz}$ followed by relaxation.
Bottom: the time-dependent plasma frequency $\omega_p(t)$.
(b)~Measured (left) and reconstructed (right) complex optical conductivity
$\Delta\sigma(\omega, t_{pp})$ plotted in the $\omega$-$t_{pp}$ space; top panels are for the real conductivity and the bottom for the imaginary.
(c)~Training loss as a function of iteration. The first 3000
iterations use the Adam optimizer, after which the algorithm
switches to L-BFGS for fine convergence of the
autodifferentiation retrieval.
}
\label{fig:experiment}
\end{figure}

\textcolor{black}{As a real test, we apply the AD method to experimental results on WTe$_2$ from OPTP measurement at
$80\,\mathrm{K}$, a  type-II Weyl semimetal in which a $T_d\to 1T'$ structural transition can be induced by pulsed laser excitation~\cite{Homes2015,Zhou2025}}. At THz frequencies, the structural transition is concomitant with a pronounced electronic percolation between the topological non-trivial and trivial states. The dataset comprises \textcolor{black}{a total of 50 pump-probe delays, i.e. $N_{t_{pp}}=50$}. Since the probe pulse duration
($\sim\!0.25\,\mathrm{ps}$) is short relative to scan length
($\sim\!9.8\,\mathrm{ps}$), we adopt a per-delay constant Drude model with smoothness regularization along the delay axis.
\textcolor{black}{The retrieved scattering rate $\gamma(t_{pp})$ [Fig.~\ref{fig:experiment}(a)] rises rapidly from approximately 4.9 THz
in the equilibrium to a peak value of 5.4 THz, followed by biexponential recovery in 7 ps, a timescale consistent with energized charge carriers and photoinduced structural dynamics. The plasma frequency $\omega_{p}$ shows a concomitant transient increase, owing to the photo-injected free carriers. The reconstructed differential optical conductivity $\Delta\sigma(\omega,t_{pp})$ closely matches the measurement at all pump-probe delays }[Fig.~\ref{fig:experiment}(b)], with residuals limited by experimental noise. 

\section{Conclusion}

In summary, we have developed a full-waveform inversion framework enabled by the automatic differentiation technique for the purpose of retrieving sub-pulse-width dynamics from two-dimensional OPTP datasets. This exceptional temporal resolution is inherently determined on the probing side only by the gate pulse rather than the much longer THz probe pulse. By structuring the complete forward computation as a differentiable program, this approach bypasses analytical gradient derivations and may be impartially applicable to systems intelligible within the framework of Drude or Lorentz model, or both, as verified numerically and experimentally above. 
Crucially, as the methodology relies only on the differentiability of the forward physical model, it could also be extended to the study of highly nonlinear and strongly coupled systems, such as those of strong electron-phonon interactions and polaronic quasi-particles, within the same unified optimization framework.

\section{Acknowledgements}
We thank Mr. Yuntian Han, for his assistance with code debugging and useful discussions. This research was supported by the Hong Kong Research Grants Council (GRF16303721, GRF16306522, GRF16307124, N$\_$HKUST631/21); the National Natural Science Foundation of China (Grant No.12122416); the National Key Research and Development Program of China (2020YFA0309603); HKUST UROP Support Grant (UROP21SC05, UROP21SC06, UROP23SC11).

\bibliography{references}

\end{document}